\documentstyle[12pt]{article}
\begin{document}
\begin{center}
{\bf ``Renormalization" Of Non Renormalizable Theories}

\medskip

     J.Gegelia, G.Japaridze, K.Turashvili and N.Kiknadze \\

\medskip

{\it High Energy Physics Institute, University street 9,
Tbilisi 380086, Georgia.}
\end{center}

\medskip

\begin{abstract}
A perturbative approach for non renormalizable theories is developed.
It is shown that the introduction of an extra expansion parameter allows
one to get rid of divergences and express physical quantities as series
with finite coefficients. The method is demonstrated on the example of
massive non abelian field coupled to a fermion field.
\end{abstract}

The presence of  divergences is one of the basic problems
of quantum field theory (QFT). The renormalization procedure handles
these divergences only for a class of theories i.e renormalizable
ones.
It is not {\it a priory} clear that nonrenormalizable theories
lack physical significance, and  moreover, in spite of the fact that most of
the fundamental interactions are described by renormalizable QFT-s, the
problem of the quantum gravitation is still open --- while Einstein's
classical theory of gravitation has substantional success, the
corresponding quantum theory is non renormalizable.

We share the opinion that the renormalizability is just a technical
requirement and has nothing to do with the physical content of
the QFT. A lot of people believe that in meaningful theories
divergences arise due to the use of a perturbative expansion,
and has been noted in various papers and various contexts [1],[2].
Of course not all of the non renormalizable theories are meaningful,
which is also true for the renormalizable ones too.
E.g. the scalar $\phi^3$ theory is renormalizable for space-time
dimensions up to six [3], but has an energy spectrum unbounded from below. On
the other hand there exist non renormalizable theories which can be
handled in some other approach (e.g. the four-fermion interaction in
$2+1$ dimensions is non renormalizable if the conventional
renormalization procedure is applied, but can be renormalized after
performing a $1/N$ expansion with $N$ being the number of
flavours [4]).

Below we are going to present a method of extracting physical
information out of the perturbative series of non renormalizable
theories. For renormalizable ones it just coincides with
the usual renormalization procedure and in that case alone can it be
interpreted in terms of counter terms. We emphasize that described
technique is unambigous.

Let us use the example of $SU(N)$ massive gauge field coupled to
fermions to illustrate how the method works. This theory, being massive,
suffers only from ultraviolet divergences, and is given by the Lagrangian:

\begin{equation}
L=-\frac{1}{4}G^a_{\mu\nu}G^{a\mu\nu}+i\bar\psi\hat D\psi-m_0\bar\psi
\psi+M_0^2A^{a\mu}A^a_\mu
\end{equation}                                       

where $G^a_{\mu\nu}$ is the gauge field strength tensor ($a=1,...N$)
and $\hat D$ is the covariant derivative with coupling $g_0$.
This theory is not renormalizable [5], and
we will work in terms of bare parameters using dimensional regularization
($n\equiv 4+2\epsilon$).
The Feynman rules for this model may be derived in the standard
manner.
Simple power counting shows that the result of any
diagram can be written in the following form:
\begin{equation}
\sum
C_{ij}(\epsilon)\left(\frac{g_0^2}{\epsilon}\right)^ig_0^j
\label{i}
\end{equation}                                       
where the coefficients $C_{ij}(\epsilon)$ are expandable as positive
power series of $\epsilon$.

Let us proceed along the lines of the usual renormalization
procedure.
To make our method more transparent we avoid any
numerical results of calculations.

We define the physical masses of vector meson and fermion as
the pole positions of their propagators and express
$m_0$ and $M_0$ in terms of physical masses $m$ and $M$ (mass
renormalization).
The wave function renormalization constants are defined
as residues at the poles.

The amputated Green's function
$<0|T(\bar\psi A^a_\mu\psi)|0>$ after mass and wave function
renormalization takes the form (Fermion legs are on mass shell
and $q$ is a momentum of the vector field):
\begin{equation}
i\Gamma_{ij,\mu}^a=it^a_{ij}\left[A(q^2)\gamma_{\mu}+
B(q^2)\sigma_{\mu\nu}q^{\nu}\right]
\end{equation}                                             

where
\begin{equation}
A(q^2)=a_0g_0+g_0^3\left[{a_1(q^2)\over \epsilon}+a_2(q^2)\right]
+...
\end{equation}

\begin{equation}
B(q^2)=b_0(q^2)g_0^3+g_0^5\left[{b_1(q^2)\over\epsilon}+b_2(q^2)
\right]+g_0^7\left[{b_3(q^2)\over \epsilon^2}+{b_4(q^2)\over \epsilon}+
b_5(q^2)\right]+...
\label{B}
\end{equation}
Note that the $\epsilon$-dependence not shown explicitly
is regular.
 (Evidently, $A$ and $B$ depend on all parameters of the theory).
$t^a_{ij}$ denote the group generators.

We introduce renormalized coupling constant as:
\begin{equation}
g=A(Q^2)/a_0
\label{g}
\end{equation}
Here $Q^2$ is the normalization point. Expressing $g_0$ from (\ref{g}) we
will have:
\begin{equation}
g_0=g-g^3\left[{a_1(Q^2)\over a_0\epsilon}+a_2(Q^2)/a_0\right]+...
\label{g_0}
\end{equation}
Substituting (\ref{g_0}) into (\ref{B}) we get:
\begin{equation}
B(q^2)=b_0(q^2)g^3+g^5{B_1(q^2,Q^2)\over \epsilon}+g^5B_2(q^2,Q^2)+
g^7\left[{B_3(q^2,Q^2) \over \epsilon^2}+
{B_4(q^2,Q^2)\over \epsilon}+B_5(q^2,Q^2)\right]+...
\label{*}
\end{equation}

Let us introduce a "related" function
$$
B^*(q^2)=b_0(q^2)g^3+g^3x^2{B_1(q^2,Q^2)\over \epsilon}+
g^5B_2(q^2,Q^2)+g^3x^4{B_3(q^2,Q^2)\over \epsilon^2}+
$$
\begin{equation}
+g^5x^2{B_4(q^2,Q^2)\over \epsilon}+g^7B_5(q^2,Q^2)+...
\label{B^*}
\end{equation}
In (\ref{*}) every inverse power of $\epsilon$ is accompanied by a $g^2$.
One can rewrite (\ref{*}) as series of $g$ and ${g^2\over \epsilon}$.
(\ref{B^*}) has been obtained by replacing this power of $g^2$ with $x^2$.
Evidently, substituting $x=g$ into (\ref{B^*}) we recover (\ref{*}).

Extracting $x^2$ iteratively from (\ref{B^*}) ($x^2$ is extracted at the
point
$q^2=Q^2$, although we could take any other normalization point) we get:
\begin{equation}
x^2=\epsilon\left[\alpha-{B_4(Q^2,Q^2)\over B_1(Q^2,Q^2)}\alpha g^2-
{B_3(Q^2,Q^2)\over B_1(Q^2,Q^2)}\alpha^2+...\right]
\label{x}
\end{equation}
Here $\alpha$ is defined by the expression:
\begin{equation}
\alpha(Q^2)\equiv {B^*(Q^2)-b_0(Q^2)g^3-g^5B_2(Q^2,Q^2)-g^7B_5(Q^2,Q^2)-...
\over B_1(Q^2,Q^2)g^3}
\end{equation}

{}From (\ref{i}) it follows that after renormalization of the wave function
and masses and substituting (\ref{g_0}) for $g_0$ any physical cross
section takes the form:
\begin{equation}
\sigma_i=\sum C^i_{ml}\left({g^2\over \epsilon}\right)^mg^l
\end{equation}
Now consider a particular physical process $ff\to ff$. The cross section has
the form:
\begin{equation}
\sigma=s_0g^4+g^6\left[{s_1\over\epsilon}+s_2\right]+...
\label{sigma}
\end{equation}
Following previously define a ``related" function
\begin{equation}
\sigma^*=s_og^4+g^4x^2{s_1\over \epsilon}+g^6s_2+...
\label{sigma*}
\end{equation}

Substitution of equation (\ref{x}) into (\ref{sigma*}) gives:
\begin{equation}
\sigma^*=s_0g^4+s_1g^4\alpha+s_2g^6+...
\label{888}
\end{equation}

The $\epsilon\to 0$ limit is non divergent for (\ref{888}), and in fact
it is an expression for $\sigma$. So, for $\sigma$ we get a
finite series in terms of $g$ and $\alpha$. Analogously we produce finite
expressions for all physical quantities.

To better understand the approach let us consider
one simple example:

Suppose we have two functions $f_1$ and $f_2$ each given by series with
divergent coefficients (in the $\epsilon\to 0$ limit):
\begin{equation}
f_1=-{g^3\over \epsilon}+{g^5\over \epsilon}+{1\over 2}{g^5\over
\epsilon^2}+...
\label{u}
\end{equation}
\begin{equation}
f_2=1+g+{g^2\over \epsilon}-{g^4\over \epsilon}+...
\label{uu}
\end{equation}
Note that $k$-th inverse power of $\epsilon$ goes together with at least
the $k$-th power of $g^2$. We again define ``related" functions (in each
term containing $\epsilon^{-k}$, $g^{2k}$ is replaced by $x^{2k}$), so:
$$
f_1^*\equiv -g{x^2\over \epsilon}+g^3{x^2\over \epsilon}+{g\over 2}{x^4\over
\epsilon^2}+...
$$
\begin{equation}
f_2^*\equiv 1+g+{x^2\over \epsilon}-g^2{x^2\over \epsilon}+...
\label{f2}
\end{equation}
Express $x^2$ iteratively from (\ref{f2}) as a power series of $g$ and
$\alpha^*\equiv f_2^*-1-g$ and substitute it into the expression of $f_1^*$.
It is easy to see that the divergences disappear. We get:
$$
x^2=\epsilon(\alpha^* +\alpha^* g^2+...)
$$
\begin{equation}
f_1^*=-(g\alpha^*-{g\over 2}\alpha^{*2}+...)
\label{ii}
\end{equation}
The right hand side of (\ref{ii}) is the expansion of
\begin{equation}
f_1^*=-glog(1+\alpha^*)=-glog(f_2^*-g)
\label{iii}
\end{equation}
Note that the same relation exist between $f_1$ and $f_2$.
Indeed we have obtained (\ref{u}) and (\ref{uu}) by ``regularizing" and
expanding the following functions:
\begin{equation}
f_1(g)=glogg^2\to glog{{g^4\over \epsilon}+1\over {g^2\over \epsilon}+1}
\label{ex1}
\end{equation}
\begin{equation}
f_2(g)=g+{1\over g^2}\to g+{{g^2\over \epsilon}+1\over {g^4\over
\epsilon}+1}
\label{ex2}
\end{equation}
So we have recovered the existing relation (\ref{iii}) between $f_1$
and $f_2$ starting from series with divergent coefficients.

We would like to note that although initially in (\ref{ex1}) and (\ref{ex2})
we had a dependence on one parameter $g$, the expansion with finite
coefficients became possible only after the introduction of an extra (not
independent) parameter $\alpha$.

So, for our example of massive vector field coupled to the fermion field, we
have expressed physical quantities in terms of
{\it two finite} parameters as series with finite coefficients. Before
concluding that these series have any status one has to show that they are
at least asymptotical. The situation is quite analogous to
conventional renormalization procedure, where  renormalizability of the
theory
does not mean that the theory is consistent. One should investigate
whether the final series are (at least) asymptotical.

Although we have not investigated the problem of consistency for our
example of a vector field coupled to fermionic one, it serves at least as an
illustration of the suggested method as
well as $\phi^4$ theory for conventional
renormalization.

To summarize, let us recall the steps we have made:
First of all the usual mass and wave function renormalizations were
performed. Next, the bare coupling $g_0$ was expressed in terms of
an effective
coupling constant $g$. Now, if we substitute $g_0$ expressed by
$g$ into physical quantities and the divergences disappear we
know that the theory is renormalizable. In the non renormalizable case
the divergences still survive and enter only via the
combination $g^\beta/\epsilon$ (where $\beta$ is fixed for given
theory and can be calculated by simple power counting). So we can
consider $g^\beta$ as an independent parameter (in fact for
mathematical rigour we have introduced ``related" functions by
replacing $g^\beta$ with $x^\beta$)
and express it as power series in $g$ and another (finite)
effective `coupling constant'. All physical quantities are expressed in
these two constants as series with
finite coefficients. Of course the validity of this series will depend
on the theory under consideration.
The suggested method coincides with ordinary renormalization
for renormalizable theories and involve the introduction of an
extra effective parameter for non renormalizable ones.

The proposed
method is easily applied within the framework of dimensional
regularization, and it is not difficult to check that the method may be
applied with other regularization schemes.

Finally we want to point out that we have applied our method to
Einstein's theory of gravitation coupled to matter fields, and detail
 this most interesting case in a separate paper.
 \\
Acknowledgements: We would like to thank  G.Jorjadze and D.Broadhurst for
useful discussions.

\end{document}